\documentclass[conference]{IEEEtran}

\usepackage{booktabs} % For formal tables
\usepackage{endnotes} 
\usepackage{listings} 
\usepackage{color}
\PassOptionsToPackage{hyphens}{url}\usepackage[bookmarks=false]{hyperref}
\usepackage[colorinlistoftodos]{todonotes}
\usepackage{tabularx}
\usepackage{xspace}
\usepackage{pifont}
\usepackage{subcaption}
\usepackage{caption}
\usepackage{hypcap}
\usepackage{listings}
\usepackage{multirow}

\newcounter{tmp}

\newsavebox{\imagebox}

\newcommand{\cmark}{\ding{51}}%
\newcommand{\xmark}{\ding{55}}%

\newcommand{\sysname}{Shadesmar\xspace} 
\newcommand{\spec}{SPEC CPU2006\xspace}
\newcommand{\mirror}{parallel\xspace} %the parallel command was taken
\newcommand{\Mirror}{Parallel\xspace} %the parallel command was taken
\newcommand{\compact}{compact\xspace}
\newcommand{\Compact}{Compact\xspace}

\newcommand{\shepherd}[1]{#1}

\definecolor{mygreen}{rgb}{0,0.6,0} 
\definecolor{mygray}{rgb}{0.5,0.5,0.5}
\definecolor{mymauve}{rgb}{0.58,0,0.82} 
\author{%{Anonymous Submission \#667}}
\IEEEauthorblockN{Nathan Burow}
\IEEEauthorblockA{Purdue University}
\and
\IEEEauthorblockN{Xinping Zhang}
\IEEEauthorblockA{Purdue University}
\and
\IEEEauthorblockN{Mathias Payer}
\IEEEauthorblockA{EPFL}
}

\usepackage[normalem]{ulem}

\begin{document}

\title{\Large \bf SoK: Shining Light on Shadow Stacks}

\maketitle

\begin{abstract}
Control-Flow Hijacking attacks are the dominant attack vector against C/C++
programs. Control-Flow Integrity (CFI) solutions mitigate these attacks on the
forward edge, i.e., indirect calls through function pointers and virtual calls.
Protecting the backward edge is left to stack canaries, which are easily
bypassed through information leaks. Shadow Stacks are a fully precise mechanism
for protecting backwards edges, and should be deployed with CFI mitigations. 

We present a comprehensive analysis of all possible shadow stack mechanisms
along three axes: performance, compatibility, and security.  For performance
comparisons we use \spec, while security and compatibility are qualitatively
analyzed. Based on our study, we renew calls for a shadow stack design that
leverages a dedicated register, resulting in low performance overhead, and
minimal memory overhead, but sacrifices compatibility.  We present case studies
of our implementation of such a design, \sysname, on Phoronix and Apache to
demonstrate the feasibility of dedicating a general purpose register to a
security monitor on modern architectures, and \sysname's deployability. Our
comprehensive analysis, including detailed case studies for our novel design,
allows compiler designers and practitioners to select the correct shadow stack
design for different usage scenarios.  

Shadow stacks belong to the class of defense mechanisms that require metadata
about the program's state to enforce their defense policies. Protecting this
metadata for deployed mitigations requires in-process isolation of a segment of
the virtual address space. Prior work on defenses in this class has relied on
information hiding to protect metadata. We show that stronger guarantees are
possible by repurposing two new Intel x86 extensions for memory protection
(MPX), and page table control (MPK). Building on our isolation efforts with MPX
and MPK, we present the design requirements for a dedicated hardware mechanism
to support intra-process memory isolation, and discuss how such a mechanism can
empower the next wave of highly precise software security mitigations that rely
on partially isolated information in a process.

\end{abstract}

\IEEEpeerreviewmaketitle

%%%%%%%%%%%%%%%%%%%%%%
\section{Introduction}
%%%%%%%%%%%%%%%%%%%%%%

Arbitrary code execution exploits give an attacker fine-grained control over
a system. Such exploits leverage software bugs to corrupt code pointers
to hijack the control-flow of an application.  Code pointers can be divided into
two categories: \emph{backward edge}, i.e., return addresses or \emph{forward
edge} pointers, such as function pointers or virtual table pointers.
Control-Flow Integrity (CFI)~\cite{abadi05ccs, burow2017control} protects
forward edges, and is being deployed by Google~\cite{llvmcfi} to protect Chrome
and Android, and Microsoft~\cite{cfguard} to protect Windows 10 and Edge. CFI
assumes that backward edges are protected. However, stack
canaries~\cite{stackguard} and safe stacks~\cite{cpi, xu2002architecture} are
the strongest backward edge protections available in mainline compilers, and
both are easily bypassed by information leaks.

Control-flow hijacking attacks that target backward edges~\cite{smashing,
ret2libc, shacham2007geometry}, e.g., Return Oriented Programming
(ROP)~\cite{shacham2007geometry, roemer2012rop, checkoway2010return}, are a
significant problem in practice, and will only increase in frequency. In the
last year, Google's Project Zero has published exploits against Android
libraries, trusted execution environments, and Windows device
drivers~\cite{pz1,pz2,pz3,pz4,pz5}. These exploits use arbitrary write
primitives to overwrite return addresses, leading to privilege escalation in the
form of arbitrary execution in user space or root privileges.  The widespread
adoption of CFI increases the difficulty for attacks on forward edge code
pointers. Consequently, attackers will increasingly focus on the easier target,
backward edges. 

C / C++ applications are fundamentally vulnerable to ROP style attacks for two
reasons: (i) the languages provide neither memory nor type safety, and (ii) the
implementation of the call-return abstraction relies on storing values in
writeable memory. In the absence of memory or type safety, an attacker may
corrupt \emph{any} memory location that is writeable. Consider, for the sake of
exposition, x86\_64 machine code where the call-return abstraction is
implemented by pushing the address of the next instruction in the caller
function, i.e., the return address, onto the stack; the callee function then
pops this address off the stack and sets the instruction pointer to that value
to perform a return. As C / C++ are memory unsafe, attackers may modify return
addresses on the stack to arbitrary values and perform code-reuse attacks such
as ROP. 

Mitigating ROP attacks requires guaranteeing the integrity of the return address
used to reset the instruction pointer after a function executes.  There are
four principle attempts to do this: (i) stack canaries, (ii) back edge
CFI, (iii) safe stacks, and (iv) shadow stacks.  Stack
Canaries~\cite{stackguard} protect against sequential overwrites of a return
address through, e.g., buffer overflows by inserting a magic value onto the
stack after the return address, which is then checked before returns.  However,
canaries are not effective against arbitrary writes where, e.g., an attacker
controls a pointer and can precisely overwrite memory.  CFI computes a valid set
of targets for indirect control-flow transfers, for returns this means any
potential call site of the function.  As shown by Control-Flow
Bending~\cite{bending}, this is too imprecise to prevent control-flow hijacking
attacks in the general case.  Safe Stacks~\cite{cpi} move potentially unsafe
stack variables to a separate stack, thereby protecting return addresses.
However, Safe Stacks offer limited compatibility with unprotected code, so are 
unlikely to be deployed.

Shadow stacks~\cite{dang2015shadow, chiueh2001rad, davi2011ropdefender,
xu2002architecture} enforce stack integrity, protecting against stack pivot
attacks and overwriting return addresses. Shadow stacks store the return address
in a separate, isolated region of memory that is not accessible by the attacker.
Upon returning, the integrity of the program return address is checked against
the protected copy on the shadow stack. By protecting return addresses, shadow
stacks enforce a one to one mapping between calls and returns, thereby
preventing ROP. Two shadow stack designs have been proposed: \compact shadow
stacks~\cite{chiueh2001rad}, which rely on a separate shadow stack pointer, and
\mirror shadow stacks~\cite{dang2015shadow}, which place the shadow stack at a
constant offset to the original stack.  These existing shadow stack designs
suffer from a combination of poor performance --- greater than the 5\% threshold
suggested by~\cite{eternal-war} and far more than the 2\% of LLVM-CFI, high
memory overhead, and difficulty supporting C and C++ programming paradigms such
as multi-threading and exception handling. 

To improve the state of shadow stack design, we conduct a detailed survey of the
design space. Our design study includes two novel designs for modern
platforms that rely on a dedicated register for performance. While a 2002
technical report~\cite{xu2002architecture} initially proposed leveraging a
dedicated general purpose register, we believe that such designs deserve renewed
attention on 64 bit architectures.
In total, our survey considers five shadow stack mechanisms. We
fully explore the trade-offs of these designs in terms of performance,
compatibility, and security. We consider the impact of high level design
decisions on runtime, memory overhead, and support for threading, stack
unwinding, and unprotected code. For the performance comparison we use \spec as
the standard benchmark, with qualitative arguments based on design for features
like threading that are not exercised by \spec. For security, we note that the
best shadow stack design approach, instrumenting function prologues and
epilogues, results in a time of check to time of use (TOCTTOU) window on x86.
The TOCTTOU window requires impeccable timing to be exploitable~\cite{bialek},
and we discuss design approaches to avoid it.  Further, we propose novel
optimizations for comparing the shadow and stack return addresses, improving the
performance of all shadow stack schemes by 25\%.  As LLVM~\cite{lattner2004llvm}
is in the process of developing a shadow stack
implementation~\cite{llvm-shadow}, the time is ripe for such a design survey and
optimizations to maximize impact.

Beyond the design of the shadow stacks, we analyze the options for guaranteeing
their integrity, including existing software solutions and two new ISA
extensions. Unlike CFI, which relies on immutable metadata stored on read-only
pages, shadow stacks, and other security mechanisms, require mutable metadata
that must be integrity protected. Integrity protection is accomplished by
isolating an area of the address space within a process, preventing attackers
from modifying it. We discuss the limitations of existing hardware mechanisms
for intra-process isolation, and propose a new primitive better suited for use
by software security mechanisms.  

Based on our design study, we propose \sysname, a new \compact shadow stack
mechanism that leverages a dedicated register for the shadow stack pointer and
our optimizations for comparing the program and shadow return addresses.
We present case studies of \sysname on Phoronix and Apache to highlight its
practicality and thoroughly evaluate it. We provide a detailed discussion of the
trade-offs between the different shadow stacks along the axes of performance,
security, and compatibility. We hope that our thorough evaluation will lead to
the adoption and deployment of shadow stacks in practice, closing a significant
loop-hole in modern software's protection against code-reuse attacks. \sysname,
along with ports of all prior shadow stack techniques to LLVM-7.0.0 is available
at \url{https://github.com/HexHive/ShadowStack}, to aid deployment of shadow
stacks.

We present the following contributions:
(i) Comprehensive evaluation of the shadow stack design space along the axes
      of performance, compatibility, and security;
(ii) Performance evaluation of each shadow stack design, including
      sources of overhead, and our optimizations for x86;
(iii) Comparative study of new ISA features that can be used to
      create integrity protected memory regions for any runtime mitigation, and
      a proposal for an intra-process isolation mechanism;
(iv) \sysname a register-based \compact performant, secure, and deployable
      shadow stack scheme and its evaluation.

%%%%%%%%%%%%%%%%%%%%
\section{Background}
%%%%%%%%%%%%%%%%%%%%

To enable security analysis of shadow stacks, we first establish our attacker
model.  Using this attacker model, we then discuss common attacks on the stack,
e.g., ROP, which overwrite return addresses for interested readers.
Knowledgeable readers may wish to move directly to our discussion of the shadow
stack design space in \autoref{sec:design}.

\subsection{Attacker Model}
%%%%%%%%%%%%%%%%%%%%%%%%%%%

As is standard for defenses that aim to mitigate exploits, e.g., CFI and Shadow
Stacks, rather than the underlying corruptions, e.g., memory or type safety, we 
assume an attacker with arbitrary memory read and write primitives. The attacker
uses these arbitrary reads and writes to inject her payload, and then corrupts a
code pointer to hijack the program's execution, executing her payload and
exploiting the application. The adversary is constrained only by standard
defenses: DEP~\cite{dep} and ASLR~\cite{team2003pax}.  We disable stack
canaries~\cite{stackguard} as they are strictly weaker than Shadow Stacks.  

For the final step of the attack, corrupting a code pointer, we assume that the
attacker only targets \emph{backward edges}, i.e., return addresses of
functions.  Protection for \emph{forward edges} is an orthogonal problem,
covered by defenses such as CFI~\cite{abadi05ccs, burow2017control}. Attacking
forward edge control flow is therefore out of scope for this paper. Also out of
scope are data-only attacks, i.e., attacks that do not corrupt code pointers. 

\subsection{Attacks on the Stack}
%%%%%%%%%%%%%%%%%%%%%%%%%%%%%%%%%

Attacks against stack integrity began with Aleph One's seminal work on stack
smashing~\cite{smashing}. To this day, control-flow information on the stack
remains an active battle ground in software security~\cite{eternal-war}.
Code reuse attacks such as ROP and Stack Pivots are the latest iteration of this
threat.  

ROP~\cite{shacham2007geometry} is a style of code-reuse attack that hijacks
application control flow by overwriting return addresses on the stack.
When the function returns, control is redirected to the attacker
chosen address.  Absent any hardening, return addresses on the stack can be
modified to target any executable byte in the program.  If a non-executable byte
is targeted, attempting to execute that byte will lead to a fault, terminating
the program.  In practice, attackers target so called ``gadgets'', which are
sequences of executable bytes ending in a return instruction that perform some
useful computation for the attacker.  The attacker's payload consists of a
sequence of addresses of such gadgets that combined perform the desired
computation, e.g., open a shell, or, in most real-world attacks map a memory
page as executable and writable and \texttt{memcpy} target shellcode to that
page before executing the injected shellcode. 

\begin{figure}[tb]
\includegraphics[scale=.8]{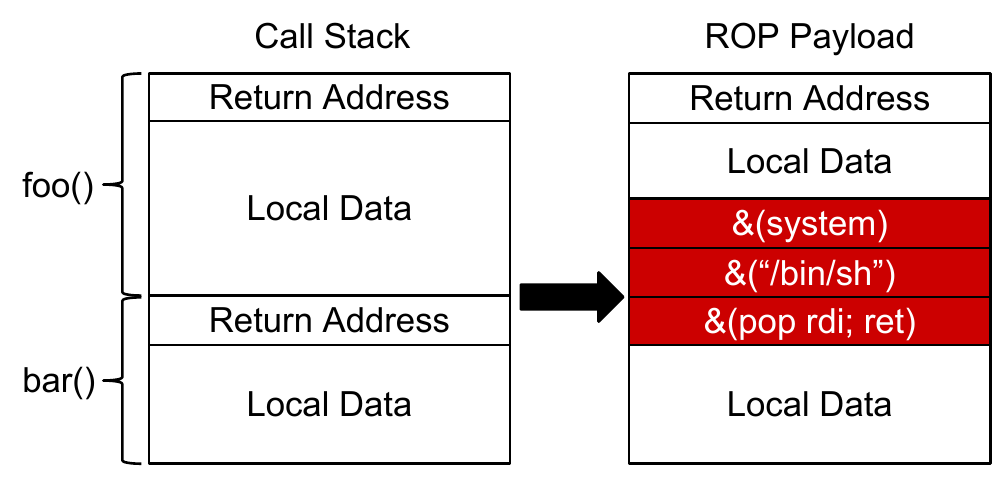}
\centering
\caption{ROP Illustration}
\label{fig:rop-example}
\end{figure}

\autoref{fig:rop-example} illustrates a payload that executes \texttt{system()}
to spawn a shell.  When function \texttt{bar()} returns, the first gadget is
executed.  Returning to the first gadget moves the stack pointer to
\texttt{\&("/bin/sh")}, which is then \texttt{pop}ped into \texttt{rdi}, and
moving the stack pointer to \texttt{\&(system)}. Consequently, the return in the
first gadget calls \texttt{system("/bin/sh")}, opening a shell.   

Stack Pivots are an attack technique wherein the adversary controls the stack
pointer, i.e., \texttt{rsp} on x86 architectures.  Consequently, instead of
having to selectively overwrite data on the stack, the attacker can move the
stack frame to a region of memory she entirely controls, thereby making, e.g.,
ROP attacks significantly easier.  This technique has also been used to bypass
ASLR~\cite{pirop, snow2013just}. While stack pivoting changes how the payload is
delivered, code-reuse attacks utilizing it must still overwrite a code pointer.
Consequently, for the purposes of shadow stacks and back edge defenses in
general, stack pivoting is just a payload delivery variant of ROP.

%%%%%%%%%%%%%%%%%%%%%%%%%%%%%%%%%%%
\section{Shadow Stack Design Space}\label{sec:design}
%%%%%%%%%%%%%%%%%%%%%%%%%%%%%%%%%%%

\begin{figure*}[t!]
\includegraphics[scale=0.8]{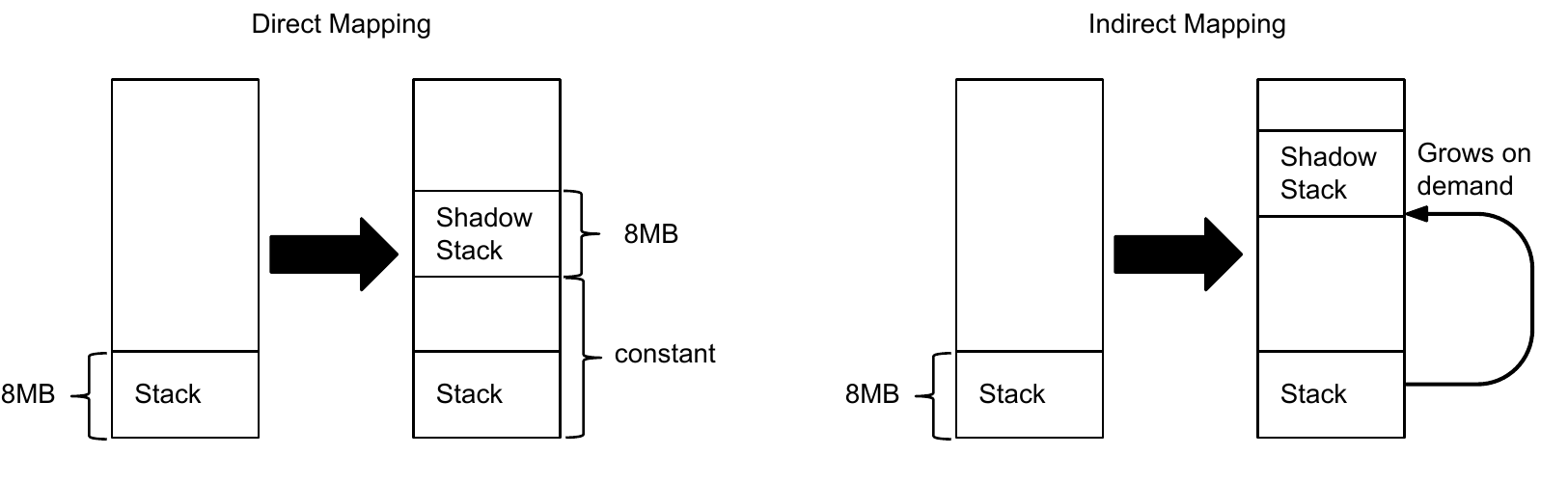}
\centering
\caption{Shadow Stack Designs -- Mapping Options}
\label{fig:shadow-mapping}
\end{figure*}

\begin{table*}[tb]
\centering
\begin{tabular}{| c | c | c | c | c | c | c |}
\hline
\multirow{2}{*}{Mapping}  & 
\multirow{2}{*}{Encoding} &
\multirow{2}{*}{Performance} &
\multirow{2}{*}{Memory} &
\multicolumn{3}{c|}{Compatibility} \\
\cline{5-7}
 & & & & Threading & Stack Unwinding & Unprotected Code \\
\hline

\multirow{3}{*}{\Compact} & Global Variable & Slow & Low 
                                       & \xmark  & \cmark & \cmark \\
\cline{2-7}
                                       & Segment         & Medium & Low
                                       & \cmark  & \cmark & \cmark \\
\cline{2-7}
                                       & Register        & Fast & Low
                                       & \cmark  & \cmark & \ding{71} \\
\hline
\multirow{2}{*}{\Mirror}  & Constant Offset & Fast & High
                                       & \xmark & \cmark & \cmark \\
\cline{2-7}
                                       & Register Offset &  Medium & High
                                       & \ding{71} & \cmark & \cmark \\
\hline
\end{tabular}
\caption{Summary of Performance Overhead, Memory Overhead, and Compatibility
  trade-offs between shadow stack mechanisms. \cmark -- supported; \xmark -- not
    supported; \ding{71} -- implementation dependent} 
\label{tbl:impl-trade-offs}
\end{table*}

For any shadow stack mechanism to be adopted in practice, it must be highly
performant, compatible with existing code, and provide meaningful security. 
We analyze the performance of each shadow stack mechanism that we identify in
terms of runtime, memory, and code size overhead qualitatively in this section,
and quantitatively in our evaluation.  Compatibility for shadow stacks means
supporting C and C++ paradigms such as multi-threading and stack unwinding, as
well as interfacing correctly with unprotected code.  Security is dictated both
by how a shadow stack mechanism validates the return address, and by any
orthogonal technique the mechanism uses to guarantee the integrity of the shadow
stack. See \autoref{sec:integrity} for details on such integrity mechanisms.  

Shadow stack mechanisms are defined by how they map from the program
stack to the shadow stack, illustrated in \autoref{fig:shadow-mapping}.  This
includes the type of mapping, as well as how the mapping is encoded in the
protected binary. We analyze five such mechanisms using the two types of shadow
stack identified by the literature: \compact~\cite{chiueh2001rad} and
\mirror~\cite{dang2015shadow}.  For \compact shadow stacks we identify three
ways to encode the mapping in the binary, and two such ways for \mirror shadow
stacks. Each of these mechanisms has unique performance and compatibility
characteristics. All shadow stack mechanisms must adopt a policy on validating
the return address.  Traditionally, this has been to compare the shadow and
program return addresses and only proceed if they match.  We examine the
security impact of utilizing the shadow return address without a comparison and
find it increases performance without impacting security. 

\subsection{Shadow Stack Mechanisms}
%%%%%%%%%%%%%%%%%%%%%%%%%%%%%%%%%%%%

Direct mappings schemes for \mirror shadow stacks use the location of the return
address on the program stack to directly find the corresponding entry on the
shadow stack.  The \mirror shadow stack is as large as the program stack, and a
simple offset maps from the program stack to the shadow stack.  Consequently,
the direct mapping trades memory overhead -- twice the stack memory usage, for
performance -- a very simple shadow stack look up.  

Indirect mapping schemes for \compact shadow stacks maintain a shadow stack
pointer, equivalent to the stack pointer used for the program stack.  The shadow
stack pointer points to the last entry on the shadow stack, exactly as the stack
pointer does for the program stack.  Maintaining a shadow stack pointer allows a
\compact shadow stack to allocate significantly less memory, as only room for
the return address is required, instead of duplicating the program stack.
Therefore, indirect mappings trade performance overhead -- from using the shadow
stack pointer, for reduced memory overhead -- by only requiring a \compact
shadow stack.  

In addition to the performance versus memory overhead trade-off, \mirror and
\compact shadow stacks have different compatibility implications.
If calls and returns were always perfectly matched, there would be no
difference.  However, the \texttt{setjmp} / \texttt{longjmp} functionality of C,
which allows jumping multiple stack frames back up the stack, and the equivalent
stack unwinding capability used by C++ for exception handling, both break the
assumption of perfectly matched calls and returns.  The direct shadow stack
paradigm naturally handles these, as C / C++ adjust the stack accordingly, and
then it uses the adjusted stack to find the appropriate shadow stack entry.  The
indirect shadow stack scheme on the other hand must know how many stack frames
the program stack has been unwound to appropriately adjust its shadow stack
pointer.  Consequently, stack unwinding leads to additional overhead for
indirect shadow stack mapping schemes, while having no effect on direct mapping
schemes.

For each shadow stack mapping scheme, there are multiple possible
mechanisms with different implications for performance and compatibility. In
particular, we introduce the use of a register for the shadow stack pointer for
\compact shadow stacks, or the offset for \mirror shadow stacks. Now that all 64
bit architectures have at least 16 general purpose registers, it is possible to
dedicate a general purpose register to the shadow stack mechanism, unlike in
2001 when the original shadow stack proposal was made~\cite{chiueh2001rad} and
only eight general purpose registers were available on x86. We find that using a
dedicated register allows \compact shadow stack mappings to be as performant as
\mirror shadow stacks, and allows \mirror shadow stacks to increase their
compatibility with multi-threading while also being more secure.

A summary of our shadow stack mechanisms and their trade-offs for each design is
shown in \autoref{tbl:impl-trade-offs}. Each row in the table represents a
shadow stack mechanism that we evaluate.  The table reports qualitative
differences between them, we refer to the evaluation in \autoref{ss:eval-shadow}
for quantitative measurements.

\subsubsection{\Mirror Shadow Stack Mechanisms}\label{ss:design-mirror}
%%%%%%%%%%%%%%%%%%%%%%%%%%%%%%%%%%%%%%%%%%%%%%%

%%%%%NHB Figure %%%%%%%%%%%
\begin{figure*}[tb]
  \begin{subfigure}[b]{0.45\textwidth}
    \begin{lstlisting}[language={[x86masm]Assembler}]
mov rax, [rsp]
mov [rsp+CONSTANT], rax \end{lstlisting}
  \caption{Constant Offset}
  \label{sf:co-impl}
  \end{subfigure}
  ~
  \begin{subfigure}[b]{0.45\textwidth}
    \begin{lstlisting}[language={[x86masm]Assembler}]
mov rax, [rsp]
mov [rsp+r15], rax \end{lstlisting}
  \caption{Offset in Register}
  \label{sf:reg-parallel-impl}
  \end{subfigure}
\caption{Direct Mapping Shadow Stack Prologues.  The epilogues execute the
inverse.}
\label{fig:direct-impl}
\end{figure*}
%%%%%%End Figure %%%%%%%%%%%%%

%%%NHB Group these in a figure as subfigures%%%%%%
\begin{figure*}[tb]
\centering
  \begin{subfigure}[b]{0.3\textwidth}
    \begin{lstlisting}[language={[x86masm]Assembler}]
mov    r10, rcx
mov	   r11, rdx
mov    rax, [rsp]
mov    rdx, GLOBAL
mov    rcx, [rdx]
mov    [rcx], rax
mov    [rcx], rsp
add    [rdx], 16
mov    rcx, r10
mov    rdx, r11 \end{lstlisting}
  \caption{Global Variable}
  \label{sf:gv-impl}
  \end{subfigure}
  ~
  \begin{subfigure}[b]{0.3\textwidth}
    \begin{lstlisting}[language={[x86masm]Assembler}]
mov rax,    [rsp]
mov r10,    gs:[0]
mov [r10],  rax
mov [r10+8], rsp
add r10,    16
mov gs:[0], r10 \end{lstlisting}
  \caption{Segment}
  \label{sf:segment-impl}
  \end{subfigure}
  ~
  \begin{subfigure}[b]{0.3\textwidth}
    \begin{lstlisting}[language={[x86masm]Assembler}]
mov rax, [rsp]
mov [r15], rax
mov [r15+8], rsp
lea r15, [r15+16] \end{lstlisting}
  \caption{Register}
  \label{sf:reg-compact-impl}
  \end{subfigure}
\caption{Indirect Mapping Shadow Stack Prologues. Note - Epilogues are the
  inverse.}
\label{fig:indirect-impl}
\end{figure*}
%%%%%NHB End figure%%%%%%%%

\Mirror shadow stack mechanisms effectively use the stack pointer as the shadow
stack pointer. The existing mechanism~\cite{dang2015shadow} places shadow stack
entries at a constant offset from the program stack. This is very efficient,
requiring no extra registers or memory access, and no instrumentation to
maintain the shadow stack pointer.  This performance benefit is offset by higher
memory overhead, compatibility problems, and lower security.  All \mirror shadow
stacks suffer from higher memory overhead, as they fundamentally require the
program stack to be duplicated. The compatibility concerns arise from requiring
a constant offset, which is limited to 32 bits for immediate operands in x86,
from the program to the shadow stack from all threads, severely constraining the
address space layout for programs with many threads, such as browsers.
Hard-coding the offset in the binary is also a security hazard, as recovering
the offset leaks the address of the shadow stack to adversaries.

To mitigate the compatibility and security concerns, we propose a new \mirror 
shadow stack mechanism. Our \mirror shadow stack mechanism encodes the offset in
a dedicated register, see \autoref{fig:direct-impl}, allowing the offset to the
shadow stack to be determined at runtime. Further, the offset may vary from
thread to thread as registers are thread local, and the offset can be set when
the thread is created.  This register is only updated once, when the offset is
determined for the thread, and therefore adds no per function call overhead
(unlike shadow stack pointers for \compact shadow stacks). 

\subsubsection{\Compact Shadow Stack Mechanisms}
%%%%%%%%%%%%%%%%%%%%%%%%%%%%%%%%%%%%%%%%

For \compact shadow stack mechanisms, the key question is where to store the
shadow stack pointer.  This decision will not impact the memory overhead of the
implementation, but does have performance and compatibility ramifications. The
shadow stack pointer will be dereferenced twice in every function: once in the
prologue to push the correct return address, and once in the epilogue to pop the
shadow return address.  Consequently, the speed of accessing the shadow stack
pointer is critical for the performance of shadow stacks that are
indirectly mapped.  There are three locations to store a
variable: in memory, in a segment, or in a register.  We discuss and evaluate
the performance and compatibility trade-offs of all three, and x86 code for each
is shown in \autoref{fig:indirect-impl}.

Using a memory location, e.g., a global variable is the simplest solution, and
we present it as a straw man.  Accessing memory is orders of magnitude
slower than accessing a value stored in a register. Even with caching, this
effect is noticeable, see \autoref{fig:design-comp}.  This slow down is
aggravated by the need for an additional move instruction to load the location
of the global variable into a register to access it -- x86 does not support 64
bit immediate values.  Further, changing memory access patterns can affect cache
behavior, with unpredictable effects on the program's performance.  An
additional problem for this scheme is that the memory must be thread local to
support multi-threaded programs.  Consequently, a scheme that has better
performance characteristics and is inherently thread local is desirable. 

Segment registers, used by existing shadow stack mechanisms~\cite{chiueh2001rad}
to store the location of the shadow stack base, are an architectural feature
left over from when physical memory was larger than the virtual address space.
Segment registers are faster to access than memory, and are inherently thread
local.  Consequently, they improve performance significantly over using a memory
location to store the shadow stack pointer, while also improving compatibility
by supporting multi-threading. We point the segment register at the base of
the shadow stack, and store the shadow stack pointer there.  Accessing the
shadow stack is thus double indirect, through the segment register and then the
shadow stack pointer.  

The earliest approach of a shadow stack scheme with a dedicated register we know
of is a 2002 technical report~\cite{xu2002architecture} that focuses on x86\_32.
We rejuvenate this idea for 64 bit architectures as general purpose registers
provide the fastest possible option for storing the shadow stack pointer.
Compared to x86\_32, the x86\_64 architecture defines twice as many general
purpose registers. The disadvantage of using a general purpose register is that
one register must be reserved for the shadow stack pointer, reducing the number
of registers available to the compiler's register allocation pass, and thereby
increasing register pressure. Increased register pressure can reduce
performance if it leads to additional register spills to the stack. Despite this
potential overhead, our evaluation finds that this is the fastest shadow stack
encoding, see \autoref{fig:design-comp}.  

\subsection{Return Address Validation}\label{ss:rav}
%%%%%%%%%%%%%%%%%%%%%%%%%%%%%%%%%%%%%%

Shadow stack mechanisms can ensure a valid return address in two ways: by either
comparing the program and shadow return addresses, or by using the shadow return
address.  Comparing the shadow and program return addresses detects corruptions
of the program return address immediately, and can halt execution.  Immediate
detection is useful during testing and debugging as it helps isolate the bug.
In deployment, however, preventing control-flow hijacking attacks only requires
that the corrupt program return address not be used.  Checking the program
return address is equivalent to a low entropy stack canary, possibly detecting
sequential buffer overflows. Consequently, the shadow stack mechanism can simply
use the return address on the shadow stack.  Doing so fully mitigates
control-flow hijacking attacks as the attacker controlled return address is not
used and avoids the overhead of comparing the return addresses.  Either policy
provides the same security: an attacker cannot control the target address of a
function return.

%%%%%%%%%%%%%%%%%%%%%%%%%%%%%%%%%%%%%%
\section{Shadow Stack Implementations}
%%%%%%%%%%%%%%%%%%%%%%%%%%%%%%%%%%%%%%

%%%%%NHB Figure %%%%%%%%%%%
\begin{figure*}[tb]
  \begin{subfigure}[b]{0.45\textwidth}
    \begin{lstlisting}[language={[x86masm]Assembler}]
pop r10
; r11 holds shadow RA
xor r11, r10
popcnt r11, r11
shl r11, 48
or r11, r10
; faults if r11 != 0
jmp r11
\end{lstlisting}
  \caption{\texttt{Fault} Epilogue}
  \label{sf:popcnt-jmp}
  \end{subfigure}
  ~
  \begin{subfigure}[b]{0.45\textwidth}
    \begin{lstlisting}[language={[x86masm]Assembler}]
pop r10
; r11 holds shadow RA
xor r11, r10
popcnt r11, r11
; will fault if r11 != 0
mov r11b, [Last_Byte_of_Page+r11] 
jmp r10 \end{lstlisting}
  \caption{\texttt{LBP} Epilogue}
  \label{sf:popcnt-mv}
  \end{subfigure}
\caption{Shadow Stack Epilogue Optimizations}
\label{fig:opt}
\end{figure*}
%%%%%%End Figure %%%%%%%%%%%%%

Each of the shadow stack mechanisms we evaluate is implemented as a backend
compiler pass in LLVM~\cite{lattner2004llvm} 7.0.0, and shares some common
implementation details. In particular, each shadow stack mechanism must
instrument calls and returns to update its shadow stack and validate the return
address before using it to transfer control. We show that the best way to
accomplish this is to instrument function prologues and epilogues. Our
implementations further include a small runtime library to set up the shadow
stacks, and support stack unwinding for \compact shadow stack schemes.
Additionally, we introduce novel peep hole optimizations for x86 epilogues.  

\subsection{Instrumented Locations}
%%%%%%%%%%%%%%%%%%%%%%%%%%%%%%%%%%%

Shadow stack mechanisms can instrument function calls either at the location of
the call instruction or in the function prologue on the callee side.  This
instrumentation is responsible for pushing the return address to the shadow
stack, and updating the shadow stack pointer for \compact shadow stacks. Returns
must be instrumented to pop from the shadow stack and validate the program
return address in the function epilogue before the control-flow transfer to
mitigate control-flow hijacking attacks.  Code that can unwind stack frames,
such as \texttt{longjmp} and C++'s exception handling mechanism, which uses
\texttt{libunwind}, must also be instrumented to maintain the shadow stack
pointer for \compact shadow stacks. Failing to handle stack unwinding correctly
can lead to false positives as the shadow and program stack are out of sync.  

The elegant solution for instrumenting calls is to place the protection in the
function prologue.  In this way, the \emph{function} is protected, not
particular call sites. The compiler does not have to distinguish between calls
to protected and unprotected functions as it would if call sites were
instrumented instead. The distinction must be made if call sites are
instrumented to keep the shadow stack in sync for \compact shadow stack where
calls and returns must be perfectly matched.  Instrumenting function prologues
and epilogues maintains this symmetry naturally, as each will be executed for
every function call. \shepherd{On x86, instrumenting the function prologues
results in a one-instruction wide Time Of Check To Time Of Use (TOCTTOU)
opportunity due to architectural limitations. The \texttt{call} instruction
pushes the return address to the stack where it may be modified by an attacker
before it is picked up by the prologue in the called function. Architectures,
such as ARM, where the address of the called function is stored in a register,
do not have this limitation.}

While the TOCTTOU window exists, given the extremely precise timing required, we
do not believe this potential weakness to be readily exploitable. Any such attack
would rely on accurately timing the victim process, and manipulating the OS
scheduler to pause the victim's execution precisely between the \texttt{call}
and \texttt{mov} instruction. After the call instruction pushes the return
address onto the stack, it remains in the cache and the \texttt{mov} instruction
can immediately use the value, resulting in a minimal window of only a few
cycles. \shepherd{Microsoft researchers proposed and redacted Return Flow Guard
(RFG) as it was vulnerable to TOCTTOU windows in the prologue and epilogue. The
Microsoft red team discovered a viable attack against their proposed epilogue,
targeting epilogues of leaf functions~\cite{bialek}.  Our proposed mechanism
halves the attack window as we jump to the verified address, but do not fully
mitigate the TOCTTOU window.  Intel Control Enforcement Technology
(CET)~\cite{intel-cet} introduces a shadow stack based on hardware and compiler
support. This extension, when available, will mitigate the TOCTTOU window on
x86 and simplify the required instrumentation.}

It is also possible to mitigate this vulnerability by using \texttt{rsp} as the
shadow stack pointer. However, doing so comes with significant side effects, see
\autoref{sec:discussion}. \shepherd{Alternatively, resolving the TOCTTOU window
requires instrumenting call sites to pass the return address in a register,
e.g., \texttt{r9} which is currently used for the sixth function argument,
changing the ABI. However, this creates compatibility problems, as protected
functions called from unprotected code would need to read the return address
from the stack. Such a scheme thus requires whole program analysis, and reduces
compatibility with unprotected code.  For users in highly sensitive environments
that are concerned about sophisticated adversaries this may be a worthwhile
trade-off.}

Our prologue and epilogue rely on the stack pointer to find the return address,
and are therefore agnostic to optimizations that delete the stack frame base
pointer.  Once our epilogue has popped the return address, we do not read it
again from memory, thereby preventing TOCTTOU attacks that modify the return
address in memory between the time it is read for the shadow stack check and the
time it is used by the return instruction.  One consequence of this is that
\texttt{ret} instructions become \texttt{pop} and \texttt{jmp} instructions.
This single transformation accounts for approximately half of the shadow stack
overhead, see \autoref{fig:ovhd-breakdown}.  Hardware solutions that avoid this
overhead are discussed in \autoref{sec:discussion}.

Stack unwinding mechanisms such as \texttt{longjmp} and C++ exceptions require
additional instrumentation for \compact shadow stacks.  \Mirror shadow stacks
are unaffected as they do not require adjustment to track stack frames, i.e.,
they do not maintain a shadow stack pointer. For \compact shadow stacks, we must
be able to unwind to the correct point on the shadow stack as well. Simply
matching return addresses does not suffice for this, as the same return address
can show up multiple times in the call stack due to, e.g., recursive calls. To
deal with this, our \compact shadow stack implementations also push the stack
pointer, i.e., \texttt{rsp}.  The stack pointer and return address uniquely
identify the stack frame to unwind to, allowing our mechanisms to support stack
unwinding.

For the shadow stack mechanisms that use a register to encode the shadow stack
mapping, ensuring compatibility with unprotected code constrains our selection
of register.  A callee saved register must be used, so that any unprotected code
that is called will restore the shadow stack pointer, but only if it is
clobbered, which helps performance.  Our implementations use \texttt{r15} in
practice. An alternative would be to use \texttt{rsp} as the shadow stack
pointer, and \texttt{r15} as the stack pointer. The ramifications of such an
implementation are considered in \autoref{sec:discussion}. 

\subsection{Runtime Support}
%%%%%%%%%%%%%%%%%%%%%%%%%%%

Our runtime library is responsible for allocating the shadow stack, and hooking
\texttt{setjmp} and \texttt{longjmp}. We add a new function in the
\texttt{pre\_init} array that initializes the shadow stack for the main program thread.
This function also initializes the shadow stack pointer for \compact
shadow stack mappings.  In particular, for segment encodings it invokes the
system call to assign the shadow stack to the segment register. \texttt{Setjmp}
and \texttt{longjmp} are redirected to versions that are aware of our shadow
stacks. These patched versions required less than 20 lines of assembly to
modify.

For \compact shadow stack mappings to support multi-threading and libunwind, we
preload a small support library.  It intercepts calls to
\texttt{pthread\_create} and \texttt{pthread\_exit} to set up and tear down
shadow stacks for additional threads. We use a patched version of libunwind, to
which we added 20 lines of code for compatibility with our shadow stacks.  These
changes are minimal, and easily deployable by having, e.g., two version of the
library on the system and a compiler flag to chose which one is linked in.
If shadow stacks are universally used to harden libraries, no such additional
support would be required. Consequently, we believe \compact shadow stacks are
readily deployable.

\subsection{Shadow Stack Epilogue Optimizations}\label{ss:opt}
%%%%%%%%%%%%%%%%%%%%%%%%%%%%%%%%%%%%%%%%%%%%%%%%

Traditionally, shadow stacks have relied on compare instructions to validate the
shadow return address and program return address are equivalent.  However, the
compare and jump paradigm is relatively expensive, potentially leading to
pipeline stalls even with branch prediction.  Consequently, as an optimization,
we explore two different methods to optimize this validation.  Our optimizations
rely on the insight that a full comparison is not required, only an equality
test.  

To replace the compare instruction of traditional shadow stack epilogues, we
propose an \texttt{xor} of the program return address and shadow return address.
This will result in \texttt{0} bits anywhere the two are identical, and
\texttt{1}s elsewhere.  x86 has an instruction, \texttt{popcnt}, that returns the
number of bits set to \texttt{1}.  Consequently, if the \texttt{popcnt} of the
\texttt{xor} of the program return address and shadow return address is
\texttt{0}, then the two are equivalent.  

We leverage the memory management unit (MMU)  to compare the \texttt{popcnt} to
zero as a side effect by creating a protection fault.  We propose two different
ways to do so: \texttt{fault} and last byte in page (\texttt{LBP}), see the code
in \autoref{fig:opt}.  For \texttt{fault}, we note that the maximum value of the
\texttt{popcnt} is 64, therefore fitting in six bits.  By shifting this value
left 48 and \texttt{or}ing it into the return address, we create a general
purpose fault for a non-canonical address form if its value is not zero, by
setting one of the high order 16 bits to one in user space. This scheme abuses
the fact that the high order 16 bits are currently unused, and may break if
those bits are utilized in future processors.  Alternately, the \texttt{LBP}
scheme creates two pages in memory, the first of which is mapped read write, the
second of which has no permissions.  We then attempt to read from the first page
at the address of the last valid byte, plus the \texttt{popcnt} value.  If the
\texttt{popcnt} value is zero, we read the last byte of the valid page,
otherwise we read from the guard page, causing the \texttt{MPU} to return a
fault.  The trade-off between the two is that the \texttt{fault} scheme requires
serialization in the processor, while the \texttt{LBP} scheme requires a memory
access and the MMU. We show the performance of both schemes in
\autoref{fig:epilogue-opt}.

%%%%%%%%%%%%%%%%%%%%%%%%%%%%%%%%%%%%%%%
\section{Hardware Integrity Mechanisms}\label{sec:integrity}
%%%%%%%%%%%%%%%%%%%%%%%%%%%%%%%%%%%%%%%

Once a shadow stack design has been chosen, the shadow stack mechanism must
guarantee the integrity of the shadow stack.  How to guarantee the integrity of
a protected region of memory is a problem faced not only by shadow stacks, but
also by all mitigations that rely on writable runtime metadata. 
Integrity guarantees are best provided by hardware solutions, though software
solutions exist and are covered here. Hardware solutions offer greater security
and performance than software solutions, and can be as generic. Hardware
solutions for integrity protecting part of the address space within a process
should be evaluated on two metrics: their performance, and the number of
supported concurrent code regions.

Existing hardware mechanisms take two different approaches to encoding access
privileges to provide integrity protection: (i) MPK which encodes access
privileges in each thread's register file, providing per thread (thread centric)
integrity, and (ii) MPX which encodes access in the individual instructions, so
that access privileges are the same across all threads and depend only on the
executed instruction (code centric). Note that thread centric solutions require
additional instructions to change the register file, consequently, code centric
solutions are (potentially) more performant as they operate in a single step,
checking an instruction's permissions, instead of first toggling bits in the
register file and then checking permissions. For code centric mechanisms, the
ability to execute the instruction grants the necessary permissions while for
thread centric mechanisms, the state of the register file determines the policy. 

Assuming code integrity and a control-flow hijacking defense such as
CFI, we prefer code centric solutions for their potential
performance and flexibility. Unfortunately, no existing code centric solution is
fully satisfactory in that they have excessive code size increases, lack
performance, and are not as flexible as required, i.e., split memory into only
two regions.  Consequently, we call for a new ISA extension that is
hardware-based for performance, supports multiple secure regions to be general
purpose (e.g., to support multiple concurrent security monitors, each with its
own protected region), and requires minimal code changes. Such an extension
would support many different security policies, as opposed to past proposals
for policy specific extensions~\cite{intel-cet, xu2002architecture, watchdog,
watchdoglite}. We show how our proposed mechanism is a code centric adaptation
of the state of the art thread centric mechanism, and thus is fully practical.  

\subsection{Thread Centric Solutions}
%%%%%%%%%%%%%%%%%%%%%%%%%%%%%%%%%%%%%

Thread centric solutions operate by changing the permissions on the pages
of the protected memory region.  Adding write permissions elevates the thread's
privileges, thereby creating a privileged region that is able to modify the
protected memory region, i.e., the shadow stack.  Removing the write permissions
ends the privileged region.  The traditional mechanism for doing this is the
\texttt{mprotect} system call.  Using \texttt{mprotect} is prohibitively
expensive as it not only requires a context switch into the kernel, but a full
page table walk to change the permissions on the indicated pages. In addition,
\texttt{mprotect} enables write capabilities for all concurrent threads and not
just for the thread writing the privileged data.

As a hardware-enforced isolation mechanism, segment registers used to provide
privilege-based isolation for x86, where the segmentation register served to
give an instruction access privileges to the protected region. For 64 bit
architectures however, x86 no longer \emph{enforces} the isolation property
while still providing the segmentation registers.

\begin{figure}[tb]
\begin{lstlisting}[language={[x86masm]Assembler}]
; Read Write (disable all MPKs) 
mov eax, 0
xor ecx, ecx
xor edx, edx
wrpkru
;protection is off, write to shadow stack
...
; Read Only (enable write disable bit for shadow stack)
mov eax, 8
xor ecx, ecx
xor edx, edx
wrpkru
\end{lstlisting}
%\vspace*{-20pt}
\caption{MPK Page Permission Toggling}
\label{fig:mpk}
\end{figure}

A new Intel ISA extension, Memory Protection Keys (MPK) aims to address this by
providing a single, unprivileged instruction that can change page access
permissions on a per-thread basis. MPK repurposes four unused bits in the page
table to assign one of sixteen keys to each page, and adds a per-thread 32-bit
register that, for each key, stores if reads or writes are disabled. The new
\texttt{wrpkru} instruction writes to this new register, selectively disabling
reads or writes for pages with a given key. This approach elegantly solves the
TOCTTOU problem of \texttt{mprotect} and allows per-thread protected regions.

The assembly to enforce privileged code regions using MPK is shown
in~\autoref{fig:mpk}.  Note that the \texttt{wrpkru} instructions requires
\texttt{edx} and \texttt{ecx} to be set to 0. Intel did not disclose why the two
registers are required to be 0, it may be for future extension of the
\texttt{wrpkru} instruction to allow a full API to be developed.  The System
V calling convention, used by Linux, uses these registers to pass the third and
fourth arguments to a function respectively.  Consequently, for functions which
take more than two arguments, it is necessary to preserve the original values of
these registers, which is accomplished by moving their values to caller save
registers, and then restoring them after the \texttt{wrpkru} instruction.
Surprisingly, this scheme is slower than MPX which must instrument almost every
memory write in the program, see \autoref{fig:integrity-ovhd} for full results.

\subsection{Code Centric Solutions}\label{ss:code-centric}
%%%%%%%%%%%%%%%%%%%%%%%%%%%%%%%%%%%

The most common code centric solution is information hiding, where a pointer
to the protected region gives any instruction access privileges. Information
hiding is attractive because it adds no additional overhead; however, it is the
weakest option as the many attacks against ASLR and other information hiding
schemes attest~\cite{gras2017aslr, goktacs2016undermining, gawlik2016enabling,
oikonomopoulos2016poking, cpi-effectiveness}.  \shepherd{Zieris and
Horsch~\cite{zieris2018leak} present a detailed study of these attacks against
shadow stacks, including proposed mitigations. Nonetheless, given the history of
successful attacks against randomization defenses, we consider information
hiding to provide minimal security for the shadow stack, and recommend against
it.}  

Software Fault Isolation (SFI)~\cite{sehr2010adapting, mccamant2006evaluating,
yee2009native} is a secure software solution for isolating intra-process address
regions.  Even the best SFI implementations~\cite{sehr2010adapting} still have
7\% overhead just for the isolation, significantly more than is acceptable in
total for a deployed security monitor.  Additionally, the x86 ISA supports an
address override prefix that limits addressable memory to 32 bits.  This can be
used to crudely separate the program's address space in a 4GB region for the
process to access, leaving all other memory for the security monitor.  4GB of
memory is insufficient for many modern applications however. Consequently, a
more flexible hardware mechanism is required.  

The Intel ISA extension Memory Protection Extension (MPX) provides a hardware
mechanism that can be used to implement segmentation~\cite{cfixx} in a flexible
manner. MPX provides a bounds checking mechanism, with four new 128 bit
registers to store the bounds, and two new primitives to perform the upper and
lower bounds checks. MPX segmentation schemes divide writes into two categories,
those that are privileged to write into the protected region, and all others.
All non-privileged writes in the code are instrumented with a bounds check to
ensure that they do not touch the privileged region. In essence, unprivileged
writes are restricted to an ``array'' of memory that consists of all unprotected
regions. This approach is surprisingly performant, see
\autoref{fig:integrity-ovhd}.

\subsection{Privileged Move}\label{ss:priv-move}
%%%%%%%%%%%%%%%%%%%%%%%%%%%%

\begin{figure}[t]
\centering
\includegraphics[scale=.7]{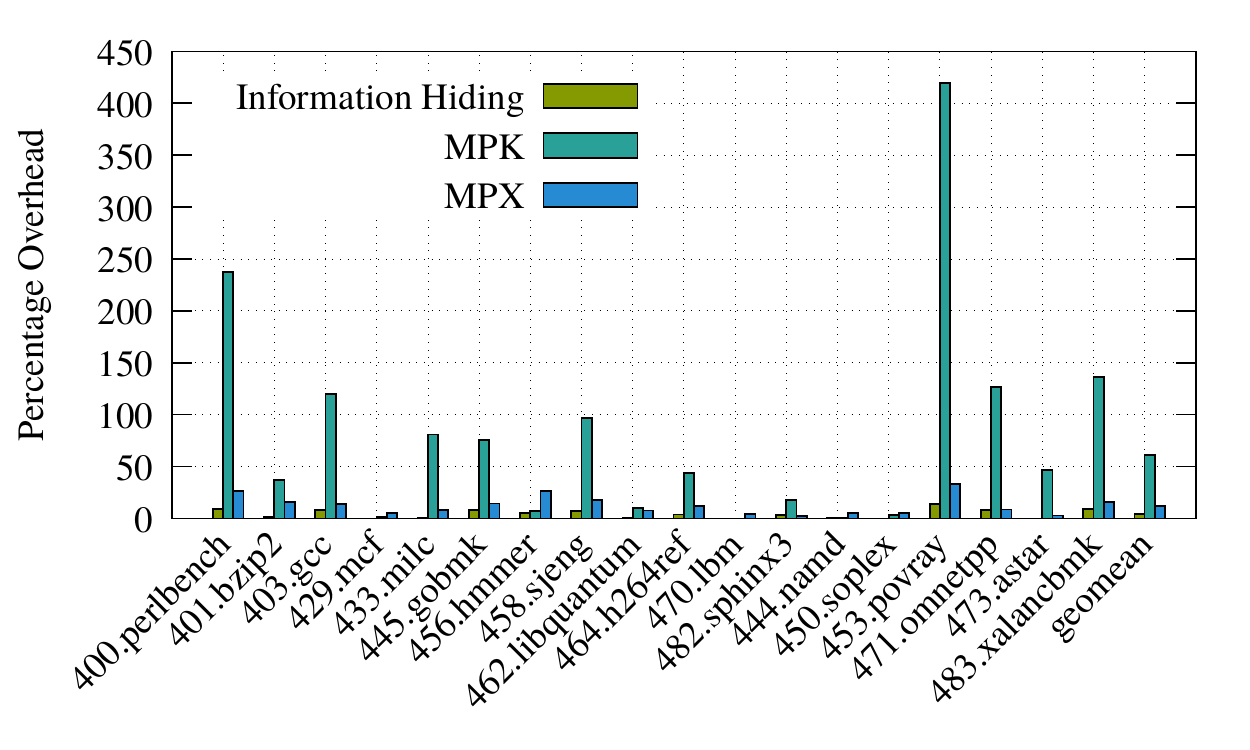}
\caption{Integrity Protection Overhead}
\label{fig:integrity-ovhd}
\end{figure}

Intel's MPK comes closest of all existing hardware mechanisms to meeting our
requirements -- it is a hardware based mechanism so should be performant, and
supports 16 code regions within a process.  However, while faster than rewriting
page tables, MPK is still too expensive to execute for every
function call, see \autoref{fig:integrity-ovhd}.  Further,
security monitors do not require a thread centric protection scheme.
Rather, a code centric scheme with a single privileged move instruction would
suffice.  This instruction could take a one byte immediate specifying the region
of memory it is allowed to write to.  Unprivileged moves would be limited by
default to the unprotected code region, allowing minimal changes.  Privileged
moves which encode their access permissions should be faster than toggling a
thread control register as MPK does.  Further, its implementation should be
largely similar, relying on the same four bits in the page table that MPK does,
and with the same checks.  The difference being that instead of referencing a
thread local state for permissions, the permissions would be encoded in the
instruction proper.

Such a privileged move instruction makes entire class of security policies
that rely on runtime metadata practical.  Currently, protecting metadata at
runtime is the bottleneck for many of these policies, covering areas as diverse
as type safety~\cite{hextype}, use after free protection~\cite{dangnull}, and
partial memory safety for function pointers~\cite{cpi}.  This hardware primitive
would allow for the creation of flexible security policies in software that can
change and adapt, such as shadow stacks.  With the availability of such a
primitive, the policies would be secure in practice, and make them deployable in
adversarial environments, instead of only being useful for testing as they
cannot withstand direct attacks.

Protection schemes that rely on new ISA extensions are unlikely to be
immediately adopted by the wider community. However, analyzing them can show
which hardware schemes are useful, hopefully paving the way for eventual broad
deployment as happened with the DEP and the NX bit.  

%%%%%%%%%%%%%%%%%%%%
\section{Discussion}\label{sec:discussion}
%%%%%%%%%%%%%%%%%%%%

Orthogonal to the main design, optimization, and protection points above there
are interesting details around dealing with unprotected code,
existing compiler optimizations with ramifications for shadow stacks, and
forthcoming hardware extensions that we include here for completeness.  

\textbf{Unprotected Code.} Unprotected code weakens the guarantees of shadow
stack schemes, as they cannot prevent a control-flow hijacking attack in the
unprotected region. Both \mirror and \compact shadow stack can be fully
compatible with unprotected code regions.  \Mirror shadow stacks are completely
oblivious to unprotected code as they do not require a shadow stack pointer.
\Compact shadow stack schemes fully support unprotected code as long as the
shadow stack pointer is not clobbered.  In particular, the register
implementation of the \compact shadow stack scheme is exposed to this. The
register implementation can handle calls into unprotected code that return
directly to protected code, as the register used is callee saved and thus
restored before protected code runs again. However, if the unprotected region
calls into protected code due to, e.g., a call back function to a sorting
routine, the shadow stack pointer may have been clobbered causing the call back
function to fail. Preventing this requires modifying the linker so that it is
aware of whether modules have been compiled with shadow stacks or not. If not,
the linker could add wrapper functions for calls across the protection boundary
that save the shadow stack pointer. With such a linker, \compact register stacks
are fully compatible with unprotected code. We leave such engineering issues as
future work.

\textbf{Tail Call Optimizations.} Tail calls allow call return pairs to be
omitted by the compiler, when, for example, a function returns the value of
another function call, or for recursive calls.  In these cases, the same program
return address can be used for the tail called function.  However, new stack
frames are required for the case where the call being optimized is the last
instruction in an arbitrary function. The optimization simply saves
instructions by omitting a call return pair by jumping directly to the callee,
which can then use one return to exit itself and the caller. As a function can
be both tail called and called normally, the full function prologue is executed
even when the function has been tail called.  To keep the shadow stack in sync,
we execute the normal shadow stack epilogue before tail calls, though we omit
the jump through the return address in these cases. Consequently, \texttt{fault}
epilogues fall back to \texttt{LBP} for tail calls, as there is no \texttt{jmp}
to modify.

\textbf{Mobile Architectures.} Beyond x86, ARM is in wide use for mobile and
embedded devices. ARM uses the \texttt{link} register to store the return
address for the current function, only pushing the return address to the stack
when additional functions are called.  Consequently, shadow stacks can
instrument function prologues without a potential TOCTTOU window. Our analysis
of the design space applies to other architectures while our epilogue
optimizations are x86 specific because of the \texttt{popcnt} instruction. Of
course, this instruction can be replaced with \texttt{shift} and \texttt{or}
instructions. We leave the evaluation of an ARM implementation as future work.

\textbf{Intel Control Enforcement Technology.} Intel has released a preview
document for a proposed new ISA extension called Control Enforcement Technology
(CET)~\cite{intel-cet}.  CET provides hardware support for shadow stacks, and
checks on forward edge indirect control-flow transfers.  CET modifies call
instructions to push the return address to a hardware protected shadow stack as
well as the program stack, and return instructions to compare the program and
shadow return addresses, raising a fault if they are not equal.  While this
technology has great promise, no release date has been made public so it is
unclear when / if it will become available. In the meantime, software solutions
for hardening programs are required. Orthogonally, other architectures and
legacy systems equally require protection.

\textbf{Tagged Architectures.} Recently, there has been renewed interest in
tagged architectures~\cite{dhawan2015architectural, darpa-ssith}, in which the
hardware associates a ``tag'' with each byte in memory that encodes a security
policy. Tags can be used to, for example, allow call instructions exclusive
rights to write to certain memory areas, preventing return addresses from being
overwritten~\cite{roessler2018protecting}. More generally, such architectures
can readily be leveraged for in-process virtual address isolation, by assigning
access permissions to each instruction in the code section, and each byte in
memory.  Such architectures are still in development however, and likely years
from widespread deployment.

\begin{figure*}[t!]
\centering
\includegraphics[scale=1.1]{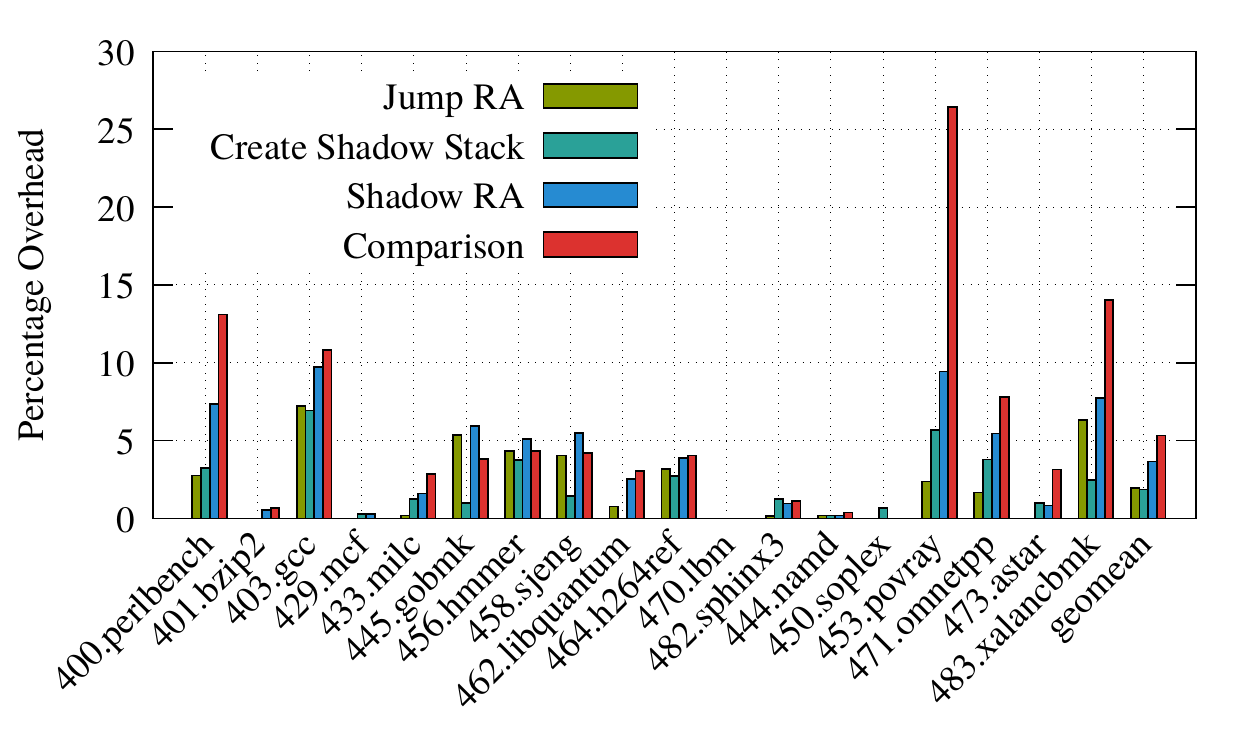}
\caption{Overhead Breakdown for \Compact Register}
\label{fig:ovhd-breakdown}
\end{figure*}

\textbf{\texttt{RSP} as Shadow Stack Pointer.} Using \texttt{rsp} as the stack
pointer instead of \texttt{r15} for \compact register based shadow stacks would
have two benefits: (i) improved performance, as shown in
\autoref{fig:ovhd-breakdown}, half the overhead of shadow stack comes from
replacing return with \texttt{pop; jmp}; and (ii) using \texttt{rsp} would
remove the TOCTTOU window as \texttt{call} instructions would directly push the
return address onto the shadow stack. However, such a design has fairly radical
compatibility implications. It would remove \texttt{push} and \texttt{pop} from
the instruction set, as these implicitly use \texttt{rsp}.  Furthermore, a
linker modification to detect unmodified code and provide wrappers for changing
the stack pointer to the expected semantics, i.e., pointing \texttt{rsp} back at
the normal stack while preserving the value of the shadow stack pointer, would
be mandatory. 
Other interesting engineering challenges include (i) support for inline
assembly, (ii) rewriting threading and standard libraries to support the two
stacks, (iii) kernel support for process creation, signal delivery, and argument
passing (which implicitly uses the rsp register), and (iv) the stack
initialization code that switches \texttt{rsp} to point to the new shadow stack.
A clean field implementation would leverage kernel support but would require
rather serious changes to the threading and standard libraries as well as the
kernel ABI.

\textbf{128 Bit Returns.} The System V ABI specifies what registers are callee
saved.  Our epilogues must preserve these registers, and of course the return
address in \texttt{rax}.  Additionally, 128 bit wide returns are possible under
the ABI and used by LLVM in practice.  These further use \texttt{rdx} for the
extra 64 bits of the return value, removing it from the pool of registers that
our epilogue can use for its computation.

\textbf{Assembly Files.} Our shadow stack mechanisms treat assembly files as
unprotected code, and do not add instrumentation. Consequently, they are allowed
to use \texttt{r15} even for the register-based shadow stack encoding.
Extending support to assembly files, including refactoring to remove uses of
\texttt{r15} is left as an engineering challenge in future work. 

\begin{figure*}[t!]
\centering
\includegraphics[scale=1.1]{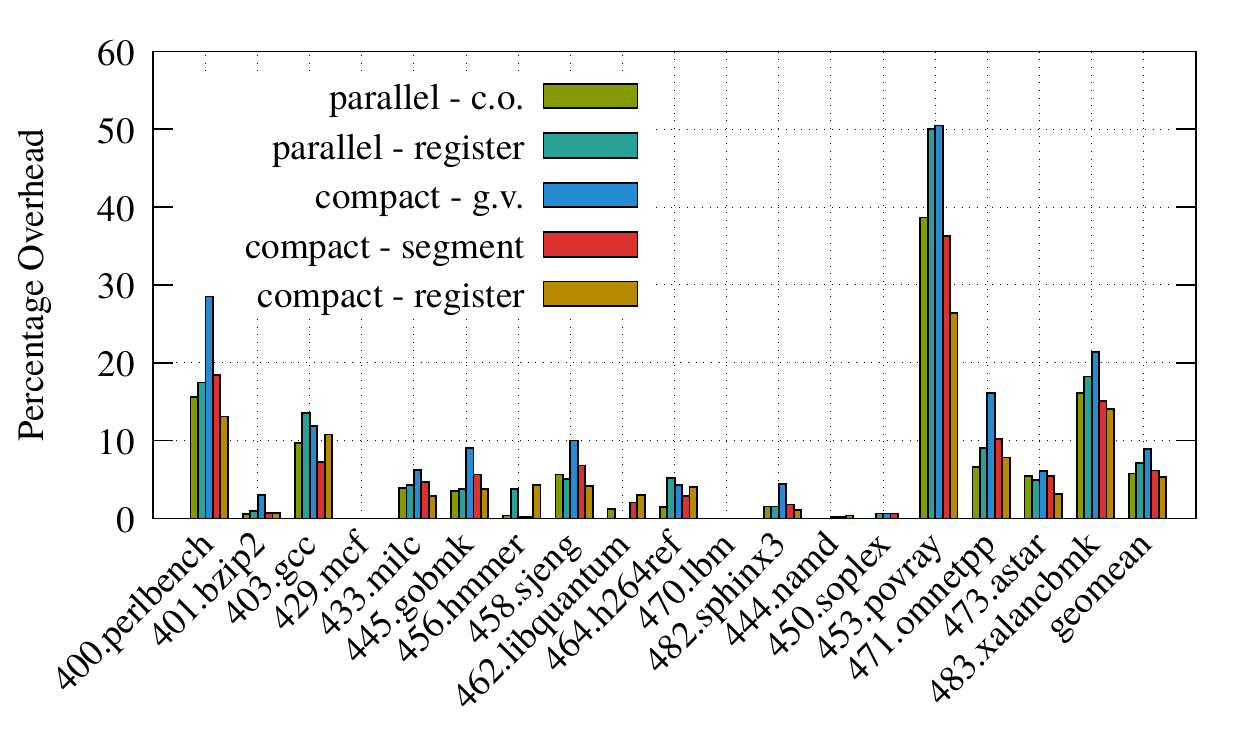}
\caption{Design Comparison}
\label{fig:design-comp}
\end{figure*}

\begin{figure}[t]
\centering
\includegraphics[scale=.7]{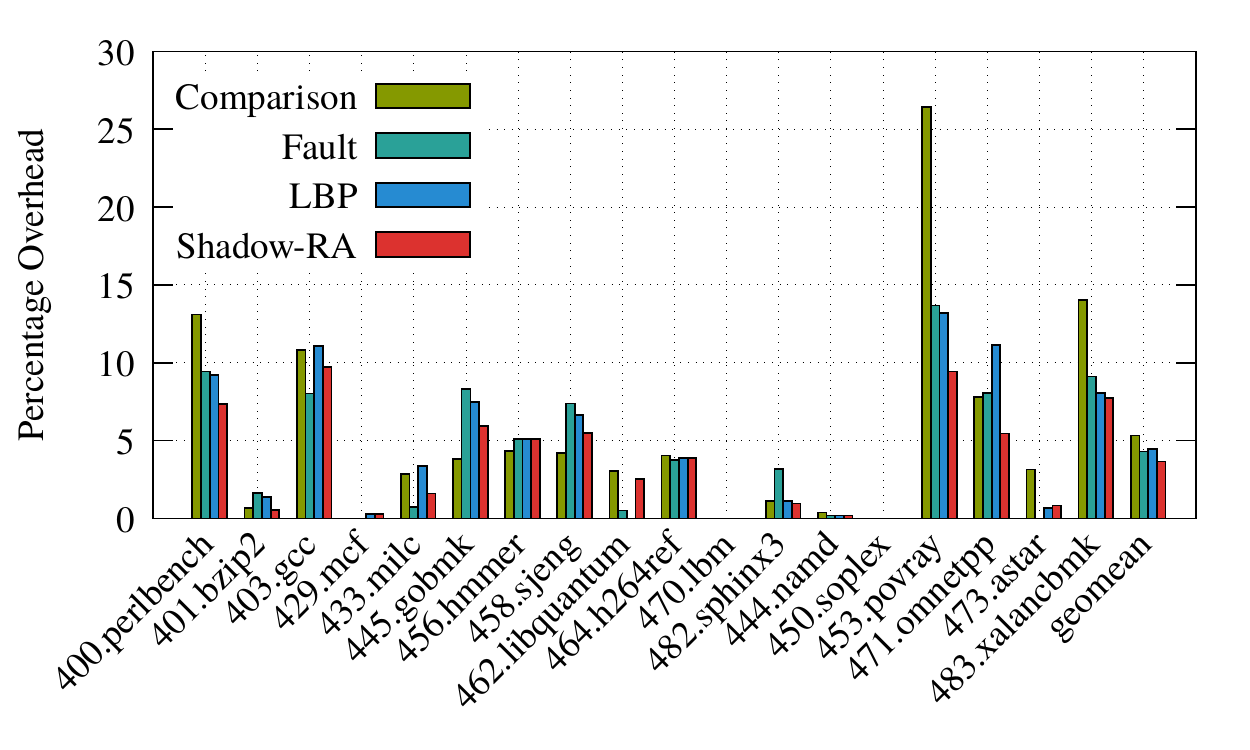}
\caption{Epilogue Micro-Optimizations}
\label{fig:epilogue-opt}
\end{figure}

%%%%%%%%%%%%%%%%%%%%
\section{Evaluation}\label{sec:eval}
%%%%%%%%%%%%%%%%%%%%

We evaluate the five different shadow stack implementations from
\autoref{tbl:impl-trade-offs}, and we examine the impact of our
proposed epilogue optimizations. Orthogonally, we evaluate the cost of providing
deterministic integrity protection for the shadow stack. Based on this
evaluation, we recommend a shadow stack mechanism, \sysname, for broad use.  To
show \sysname's practicality, we present two deployability case studies: Phoronix and the
Apache web server. The Phoronix benchmarks are common use cases for widely used,
real-world applications, and Apache is the most popular web server.
All of our evaluation is done on an Intel(R) Xeon(R) Bronze 3106 CPU at 1.7GHz,
with 48GB memory, running Debian-9.3.0. This was the first machine that
supported MPK in 2017 when this work began, and is used for all
experiments for consistency. Our results do not change on Skylake 3.0GHz
desktops with 16GB of memory running Ubuntu 16.04, which were also used during
development. We compile software at O2 and for \spec we use the default
configuration with three reportable runs on the ref dataset.

\subsection{Shadow Stack Evaluation}\label{ss:eval-shadow}
%%%%%%%%%%%%%%%%%%%%%%%%%%%%%%%%%%%%

For each of the five different shadow stack designs, we first evaluate their
performance on \spec.  For the existing shadow stack designs identified in
\autoref{sec:design}, we reimplemented them on LLVM 7.0.0 to control for
performance effects from compiler improvements. For these experiments, we used
the traditional \texttt{cmp}-based epilogue, and information hiding to protect
the shadow stack. The results are in \autoref{fig:design-comp}. Note that the
\mirror shadow stack constant offset implementation and the \compact shadow
stack register implementation are within measurement noise of each other at
5.78\% overhead and 5.33\% respectively. This removes the performance
justification for \mirror shadow stack's greater memory use, if a dedicated
register is used for the shadow stack pointer. The register based \mirror shadow
stack scheme, needed for compatibility as per \autoref{ss:design-mirror}, has 7.10\%
overhead, noticeably more than the constant offset version. Consequently, the
performance case for \compact shadow is even more compelling when equally
compatible designs are considered.  The \compact and \mirror shadow stacks have
effectively the same code size impact as well, 15.57\% and 14.88\% respectively.
Consequently, we recommend \compact shadow stacks.    

\autoref{fig:epilogue-opt} shows the overheads for the \compact shadow stack
register implementation with our different epilogue optimizations.
The traditional \texttt{cmp}-based epilogue has 5.33\% overhead, 25\% more than
our optimized epilogues at 4.31\% for the \texttt{fault} epilogue,
and 4.44\% for the \texttt{LBP} epilogue.  Further, the \texttt{cmp} epilogue
has significant outliers on perlbench, povray, and Xalancbmk.  Consequently, we
believe our epilogue optimizations are highly effective as they not only reduce
overhead but also reduce its variation.  As the \texttt{fault}-based epilogue is
faster (albeit marginally) and does not require additional changes to the
address space (\texttt{LBP} introduces guard pages), we recommend it for
vulnerability discovery settings, e.g., software testing and fuzzing. Using the
shadow return address without any comparison as discussed in \autoref{ss:rav}
results in 3.65\% overhead, and is our recommendation for deployment. 

We break down the sources of overhead within the \compact shadow stack register
implementation in \autoref{fig:ovhd-breakdown}.  Changing the \texttt{ret}
instruction to a \texttt{pop; jmp} sequence has 1.97\% overhead (the overhead is
likely due to the loss of the CPU's return value prediction).  Maintaining the
shadow stack but leaving the normal return instruction has 1.85\% overhead. If
the epilogue jumps through the shadow stack return address, there is 3.65\%
overhead, effectively the sum of the return instruction transformation and
maintaining the shadow stack, as expected.  Our experiment highlights an
opportunity for architectural improvement: moving the return stack buffer to the
shadow stack would recover most of the overhead and, due to the compact design
and fixed layout of the shadow stack, could simplify the management of that
buffer and possibly improve performance.

Our last experiment on \spec evaluates the overhead of our three different
shadow stack integrity mechanisms. For these experiments, we used a \compact
shadow stack with the register implementation and the \texttt{fault}-based
epilogue. The results are in \autoref{fig:integrity-ovhd}. As expected, the
information hiding scheme is the fastest with 4.31\% overhead.  The MPX-based,
code centric, isolation scheme was the next fastest with 12.12\% overhead on
average.  The MPK thread centric, isolation scheme had 61.18\% overhead. Our
finding is in line with Erim~\cite{vahldiek2018erim} which finds that adding a
permission switch to a direct call increases the number of cycles for the call
from 8 to 69.  Consequently, we conclude that MPK is serializing execution, and
was not intended for hot path use. MPX has a code size increase of 41.67\% vs
21.24\% for MPK. Neither the MPX nor MPK overhead numbers are acceptable for a
deployed mechanism, highlighting the need for our proposed privileged move
instruction, as per~\autoref{ss:priv-move}.

\subsection{\sysname Case Studies}
%%%%%%%%%%%%%%%%%%%%%%%%%

\begin{table}[b]
\centering
\begin{tabular}{|l|r|r|}
\hline
Benchmark & Overhead & Deviation \\
\hline
sqlite & 8.94\% &	0.22\% \\
flac & 1.19\% &	0.85\% \\
MP3 & 1.47\% &	0.28\% \\
wavpack & 0.35\% &	0.15\% \\
crafty & 0.84\% &	0.15\% \\
hmmer & 0.28\% &	0.42\% \\
LZMA & 0.84\% &	0.29\% \\
apache & -2.05\% &	0.40\% \\
minion-graceful & 1.18\% &	0.16\% \\
minion-quasigroup & 3.39\% &	0.13\% \\
\hline
\end{tabular}
\caption{Phoronix Benchmark Results}
\label{tbl:phoronix}
\end{table}

\begin{table}[tb]
\centering
\begin{tabular}{|l|r|r|r|}
\hline
\multirow{2}{*}{File Size} & \multicolumn{3}{r|}{Simultaneous Connections} \\
\cline{2-4} 
                               &  1     &  4     &  8  \\
\hline
70K - HTML                     & 6.21\% & 0.63\% & -0.40\% \\
1.4M - Image                   & 1.13\% & 0.45\% & -0.31\% \\
\hline
\end{tabular}
\caption{Apache Throughput Reduction}
\label{tbl:apache}
%\vspace{-1.5em}
\end{table}

We believe that \sysname --- a \compact, register based shadow stack that
directly uses the shadow RA, and relies on information hiding to protect the
shadow stack --- is the best candidate for adoption by mainline compilers based
on our initial experiments with \spec.  Consequently, we present a more in depth
evaluation of \sysname here, on real world applications of interest to potential
users of shadow stacks. Note that information hiding still significantly raises
the bar for attackers by requiring an information leak, and a write to a region
of memory with only one pointer into it (the shadow stack pointer) to bypass
\sysname. \sysname exclusively keeps the shadow stack pointer in a register,
making leaking the location of the shadow stack extremely difficult.
Nonetheless, from a security perspective, information hiding is fundamentally
broken as discussed in \autoref{ss:code-centric}. We recommend it only because
of the resistance to deploying any protection mechanism with greater than 5\%
overhead~\cite{eternal-war}.

To demonstrate the usefulness of \sysname for real software, we run benchmarks
from the Phoronix test suite for typical desktop user experiences, and benchmark
the throughput of the Apache webserver. For all case
studies, \sysname has minimal performance impact while greatly increasing
security by removing \emph{backward edge} control-flow transfers from the attack
surface. In particular, this shows that on modern 64 bit architectures with 16
general purpose registers, dedicating one general purpose register to a
security mechanism is acceptable in practice.

\textbf{Phoronix.} We run ten benchmarks from Phoronix with workloads including
databases, audio encoding, data compression, chess, protein sequencing, and
their version of Apache. These workloads are representative of common workloads
for user space computation. The results are in \autoref{tbl:phoronix}.
\shepherd{The only benchmark with high overhead is sqlite. Our primary source of
overhead is instrumenting calls and returns, however sqlite has the same
frequency of function calls as other benchmarks.  After further analysis, we
attribute the performance difference to code layout changes (affecting the
instruction cache) and increased register pressure.} 

For eight of the ten benchmarks, the overhead is less than 2\%; for five
benchmarks overhead is within 1\%; and it is within measurement noise of zero
for two benchmarks. Consequently, we believe that \sysname is performant enough
to be deployed in desktop computing environments, and that users would not
notice any slow down.

\textbf{Apache.} To evaluate \sysname in server settings, we
benchmarked the throughput of Apache with \sysname instrumentation.  For this
experiment, we used two different files, a 70KB HTML file and a 1.4MB image
file, representative of the average size of webpages in 2016~\cite{pagesize}.
The experiment was run on \texttt{localhost} to minimize measurement noise from
network effects. Throughput was measured over five minutes using the standard
\texttt{ab} tool. Note that the overhead drops with the number of connections,
and file size, and is non-existent for eight concurrent connections, as shown in
\autoref{tbl:apache}.  This demonstrates that \sysname has no impact on the
performance of IO bound applications like servers.

%%%%%%%%%%%%%%%%%%%%%
\section{Related Work}
%%%%%%%%%%%%%%%%%%%%%

Prior work on code-reuse attacks and defenses has focused on three major areas:
(i) offensive papers that seek to fully evaluate the potential of code-reuse
attacks, (ii) CFI defense mechanisms for mitigating forward edge attacks, and
(iii) shadow stacks for mitigating backward edge attacks.

\textbf{Code-Reuse Attack Surface.} Code-reuse attacks as an attack vector
began with the original ROP attack~\cite{shacham2007geometry}. Since then, the
research community has worked to fully understand the scope of this attack
vector.  Follow on work established that any indirect control-flow transfer
could be used for code-reuse attacks, not just
returns~\cite{checkoway2010return, jop}. JIT-ROP~\cite{snow2013just} showed how
just in time compiled code, like JavaScript, can be abused for code-reuse
attacks.  Counterfeit Object Oriented Programming (COOP)~\cite{coop} specialized
code reuse attacks for C++ programs, while PIROP~\cite{pirop} shows how to
perform ROP in the face of ASLR.  Control Jujustu~\cite{controljujutsu},
Control-Flow Bending~\cite{bending}, and Block-Oriented
Programming~\cite{ispoglou18ccs} showed that CFI defenses cannot prevent
code-reuse attacks in general. Newton~\cite{van2017dynamics} provides a
framework for analyzing code-reuse defenses' security.  \shepherd{Side channels
are a powerful primitive to attack existing code reuse defenses, e.g., by
widening TOCTTOU windows~\cite{allan16acsac} or by carefully monitoring
reads/writes~\cite{jurczyk13}.}

\textbf{Control-Flow Integrity.} CFI~\cite{abadi05ccs}
mitigates forward edge code-reuse attacks. CFI mechanisms work by using static
analysis to create an over approximation of the control-flow graph (CFG), and
then enforce at runtime that all transitions must be within the statically
computed CFG. After the initial proposal, follow on research
has removed the need for whole program analysis~\cite{mcfi, picfi}, and
specialized CFI to use additional information in C++ programs when protecting
virtual calls~\cite{vtrust, vtable-interleaving}. To improve the precision of
the CFG construction underlying CFI, more advanced static analysis techniques
have been proposed~\cite{vip}. Alternately, dynamic analysis-based approaches
that leverage execution history~\cite{patharmor}, or analyze execution history
on a separate core~\cite{pittypat} significantly increase the precision of CFI,
and thereby the security it provides. See Burow et al.~\cite{burow2017control}
for a survey of CFI techniques.  

Alternatives to CFI for forward edge protection have been proposed.  Code
Pointer Integrity (CPI)~\cite{cpi} isolates and protects code pointers, thereby
keeping them from being corrupted. CPI included a proposal for Safe Stacks which
rely on a precise escape analysis for stack variables, and other inter-procedural
analysis to divide the stack into two new stacks: a safe stack with the return
address, and variables that cannot be accessed through pointers, and an unsafe
stack.  Safe stacks have significant compatibility problems, particularly with
unprotected code and without full program analysis the conservative analysis
ends up allocating a large number of unsafe stack frames, resulting in
unnecessary overhead.  CFIXX~\cite{cfixx} provides object type integrity by
protecting the virtual table pointers of C++ objects, thereby precisely
protecting virtual dispatch.

\textbf{Shadow Stacks.} Prior work is split between binary translation
solutions~\cite{prasad2003binary, erlingsson2006xfi, davi2011ropdefender,
payer11vee, payer12oakland, qiao2015principled, payer15dimva, ge2017griffin} and
compiler-based solutions~\cite{chiueh2001rad, dang2015shadow,
quach2017supplementing, prakash2015defeating, onarlioglu2010g, zieris2018leak}.
The binary solutions employ binary rewriting to add trampolines to the
shadow stack instrumentation, and may enforce additional policies such as CFI,
or utilize Intel's Process Trace (PT) feature and an additional core to analyze
the process trace~\cite{ge2017griffin}.  Compiler-based solutions come in
three flavors: those that only attempt to prevent stack
pivots~\cite{quach2017supplementing, prakash2015defeating}, an attempt to remove
all ROP gadgets from the binary~\cite{onarlioglu2010g}, and finally full shadow
stacks~\cite{chiueh2001rad, dang2015shadow}, which offer the strongest security.
\sysname builds on full shadow stacks and introduces a dedicated shadow stack
register to improve performance for \compact shadow stacks, and compatibility by
fully supporting stack unwinding.  We also introduce hardware mechanisms to
integrity protect the shadow stack.

%%%%%%%%%%%%%%%%%%%
\section{Conclusion}
%%%%%%%%%%%%%%%%%%%

With the increasing deployment of CFI to protect against forward-edge attacks,
backward-edge defenses are required to fully mitigate control-flow hijack
attacks. 
We conduct a qualitative and quantitative study of the design
space of shadow stacks along performance, compatibility, and security
dimensions. Based on this study, we believe that \sysname, a register-based
\compact shadow stack that is compatible with all required C/C++ paradigms,
should be deployed. We provide implementations of all known shadow stack
schemes, in addition to \sysname, in LLVM 7.0.0 to aid the deployment of shadow
stacks.
Our case studies on Apache, where we had no performance impact for real work
loads, and Phoronix where we had less than 2\% overhead for 8 of the 10
benchmarks show the feasibility of using a dedicated general purpose register
for shadow stacks. 
Orthogonally, we show that currently no existing hardware mechanism is usable in
practice for intra-process address space isolation, and propose a new
code-centric mechanism to fit this need for general security monitors that
require mutable metadata.
Our design study of shadow stack demonstrates that they have low performance and
memory overhead, support all required C/C++ paradigms, and fill an important
security gap left by deployed CFI mechanisms against control-flow hijacking. 
\shepherd{The source code of our prototype implementations are available at
\url{https://github.com/HexHive/ShadowStack}.}

%%%%%%%%%%%%%%%%%%%%%%%%%%%
\section*{Acknowledgements}
%%%%%%%%%%%%%%%%%%%%%%%%%%%

\shepherd{We thank our shepherd Anders Fogh and the anonymous reviewers for
their insightful comments.  This research was supported by ONR awards
N00014-17-1-2513, by CNS-1801601, and a gift from Intel corporation. Any
opinions, findings, and conclusions or recommendations expressed in this
material are those of the authors and do not
necessarily reflect the views of our sponsors.}

\bibliographystyle{IEEEtran}
\bibliography{biblio}

\end{document}